\documentclass[12pt,preprint]{aastex}
\usepackage{emulateapj5,psfig}

\newcommand{\vlsr}	{\mbox{$ v_{\rm \scriptscriptstyle LSR} $}}
\newcommand{\vgsr}	{\mbox{$ v_{\rm \scriptscriptstyle GSR} $}}
\newcommand{\kms}       {\mbox{$~\rm km~s^{-1}$}}
\newcommand{\mdyn}	{\mbox{$ M_{\rm \scriptscriptstyle dyn} $}}
\newcommand{\mlum}	{\mbox{$ M_{\rm \scriptscriptstyle lum} $}}

\def\lesssim{\mathrel{\hbox{\rlap{\hbox{\lower4pt\hbox{$\sim$}}}\hbox{$<$}}}}
\def\gtrsim{\mathrel{\hbox{\rlap{\hbox{\lower4pt\hbox{$\sim$}}}\hbox{$>$}}}}
\newcounter{ionnum}
\newcommand{\myion}[2]	{\setcounter{ionnum}{#2}%
			 {#1}\hspace*{0.3em}%
			 {\scriptsize\Roman{ionnum}}}
\newcommand{\HI}	{\myion{H}{1}}
\newcommand{\mhi}	{\setcounter{ionnum}{1}%
			 \mbox{$ M_{\rm H\hspace*{0.1em}%
			        {\scriptscriptstyle\rm\Roman{ionnum}}}$}}

\slugcomment{\it Accepted for Publication in the Astrophysical
Journal Letters}
\shortauthors{Robishaw, Simon, \& Blitz}
\shorttitle{\HI\ Imaging of LGS~3}

\begin{document}


\title{\HI\ Imaging of LGS~3 and an Apparently Interacting High-Velocity Cloud}

\author{
Timothy Robishaw,
Joshua D.\ Simon,
and
Leo Blitz}

\affil{Department of Astronomy, University of California at Berkeley,
601 Campbell Hall, Berkeley, CA 94720-3411;}

\email{robishaw@astro.berkeley.edu, jsimon@astro.berkeley.edu, blitz@astro.berkeley.edu}


\begin{abstract}

We present a $93\arcmin~\times~93\arcmin$ map of the area near the
Local Group dwarf galaxy LGS~3, centered on an \HI\ cloud 30\arcmin\
away from the galaxy.  Previous authors associated this cloud with
LGS~3 but relied on observations made with a 36\arcmin\ beam.  Our
high-resolution (3\farcm4), wide-field Arecibo observations of the
region reveal that the \HI\ cloud is distinct from the galaxy and
suggest an interaction between the two. We point out faint emission
features in the map that may be gas that has been tidally removed from
the \HI\ cloud by LGS~3.  We also derive the rotation curve of the
cloud and find that it is in solid-body rotation out to a radius of
10\arcmin, beyond which the rotation velocity begins to decline.
Assuming a spherical geometry for the cloud, the implied mass is $2.8
\times 10^{7}\, (d/\mbox{Mpc})\ M_{\odot}$, where $d$ is the distance
in Mpc.  The observed \HI\ mass is $5.5 \times 10^6\,
(d/\mbox{Mpc})^2\ M_{\odot}$, implying that the cloud is dark-matter
dominated unless its distance is at least 1.9~Mpc.  We propose that
the cloud is a high-velocity cloud that is undergoing a tidal
interaction with LGS~3 and therefore is located roughly 700~kpc away
from the Milky Way.  The cloud then contains a total mass of
$\sim\!\!2.0 \times 10^{7} M_{\odot}$, 82\% of which consists of dark
matter.

\end{abstract}



\section{Introduction}
The biggest challenge facing studies of high-velocity clouds (HVCs) is
that their distances and masses are almost completely unknown. We
attack this problem by using the upgraded Arecibo telescope to
completely map a large area ($2.4\ \mbox{deg}^{2}$) around the Local
Group dwarf galaxy LGS~3 and a newly identified HVC.  If the HVC and
the dwarf galaxy are interacting, as we will argue, then this system
presents a unique opportunity to constrain the distance and mass of an
HVC.

The \HI\ cloud next to LGS~3 appears to have been first detected by
Hulsbosch in 1982 using the Dwingeloo 25~m telescope
\citep{christiant83}.  \citet{hulsboschw88} listed the cloud in three
entries in their HVC survey, at $(\ell,b,\vlsr\,[\kms]) =
(127\degr,-41\degr,-331)$, $(128\degr,-41\degr,-329)$, and
$(127\degr,-42\degr,-352)$, all of which they consider to be part of
LGS~3, which is located at $(\ell,b,\vlsr\,[\kms]) =
(126\fdg75,-40\fdg89,-287)$.  The original \HI\ observations of LGS~3
by \citet{thuanm79}, made using Arecibo, did not detect the cloud
because they were directed at the optical position of the galaxy
rather than 30\arcmin\ away.  \citet{christiant83} noted the clear
velocity gradient across the cloud, but they were unable to draw any
conclusions about the nature of the cloud from the
36\arcmin-resolution observations.  The cloud was also seen in the
Leiden/Dwingeloo Survey (LDS) of Galactic Neutral Hydrogen
\citep{hartmannb97} and was found by \citet{blitzr00} to be
substantially larger in both area and \HI\ mass (if at the same
distance as the galaxy) than LGS~3 itself.  \citet{blitzr00} further
noticed that the velocity of the cloud was different from that of the
dwarf galaxy by $-50\kms$ and postulated that the cloud could have
been removed from LGS~3 by ram-pressure stripping.  However, they also
argued that the HVC should have a less negative velocity than the
dwarf galaxy in the case of ram-pressure stripping, making this
interpretation doubtful.

In \S \ref{sec:obs}, we describe our observations and discuss the data
reduction.  Our analysis of the data is presented along with our map
of the HVC and its rotation curve in \S \ref{sec:results}.  In \S
\ref{sec:discussion}, we discuss our results and evidence for a
possible interaction between the HVC and LGS~3.


\begin{figure*}[!t]
\psfig{figure=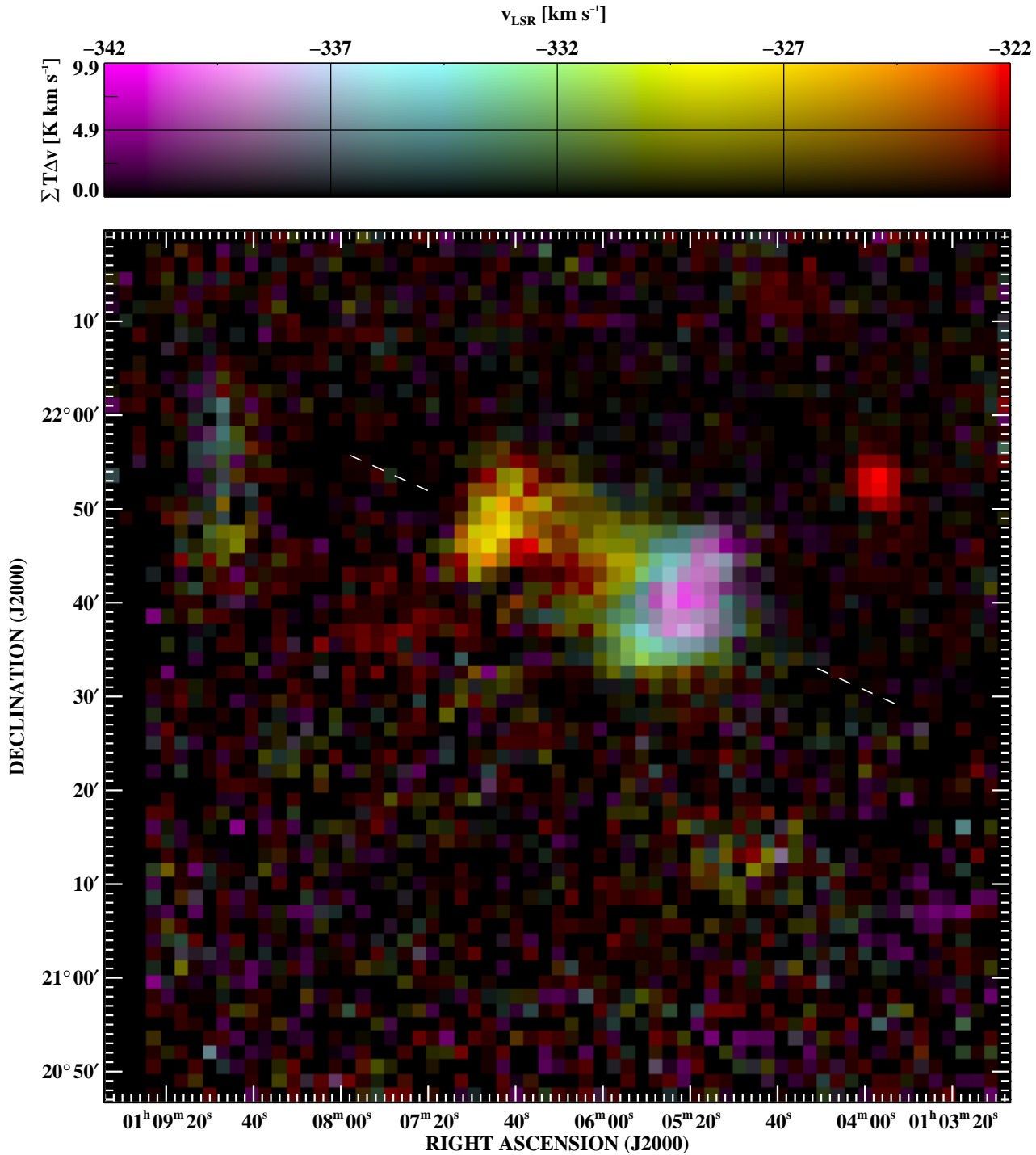}
\caption{High-resolution \HI\ map of HVC and LGS~3. This
color-intensity (velocity-column density) image consists of $\sim$10
hr of on-source integration time acquired over the course of 5
nights. The pixel size is 1\farcm5, or slightly less than half of a
beamwidth.  Visible in the map are the Local Group dwarf galaxy LGS~3
(the bright red object at $(\alpha,\delta)_{2000} =
(01^{\rm\scriptstyle h}03^{\rm\scriptstyle m}54^{\rm\scriptstyle s},
+21\degr 53\arcmin)$ in the upper right), a compact HVC (the large
double-lobed cloud in the center) with a 14\kms\ gradient across it,
and two faint features (to the left and lower right of the HVC) that
we believe are remnants of a tidal interaction between LGS~3 and the
HVC. The dashed white line shows the major axis of the HVC.
\label{fig:map}}
\end{figure*}


\section{Observations and Data Reduction}
\label{sec:obs}

The observations were conducted over five nights in 2000 November
using the upgraded Arecibo telescope.\footnote{The Arecibo Observatory
is part of the National Astronomy and Ionosphere Center, which is
operated by Cornell University under a cooperative agreement with the
National Science Foundation.} We employed the L-band narrow receiver
and the 9-level, dual-polarization correlator configuration to yield
spectra with a velocity resolution of 0.644\kms. At the frequency of
\HI\ observations, the telescope has a half-power beam width of
3\farcm4, a main beam efficiency of 0.48, and a gain of 7.2~K/Jy
\citep{heiles00}.

The data we present in this paper include 3 nights of on-the-fly (OTF)
maps, Nyquist sampled in right ascension and declination, and 2 nights
of drift scans.  The OTF maps cover a 93\arcmin\ $\times$ 93\arcmin\
region and the drift scans add sensitivity in a 40\arcmin\ $\times$
40\arcmin\ area between and including the HVC and LGS~3.  The
integration time was $\sim$22 s beam$^{-1}$ for most of the map and
$\sim$92 s beam$^{-1}$ in the area covered by the drift scans.  We
reached rms sensitivity levels at the full velocity resolution of 75
mK beam$^{-1}$ and 34 mK beam$^{-1}$ in those regions, respectively.

Each night we observed a single off-source position, which we used to
remove the bandpass shape for all of the spectra from that night.
Gain calibration was provided by the injection of a known signal into
the correlator once per strip.  We fitted and removed a linear
baseline from each spectrum. We developed a new algorithm to remove
spectral standing waves that were present in the data.  The drift
scans were resampled to match the sampling of the OTF maps, and we
then coadded each night's observations to form a single data cube.
The reduction will be described in more detail in a future paper
presenting the results of our survey of 27 HVCs and Local Group dwarf
galaxies.


\begin{figure*}[!t]
\psfig{figure=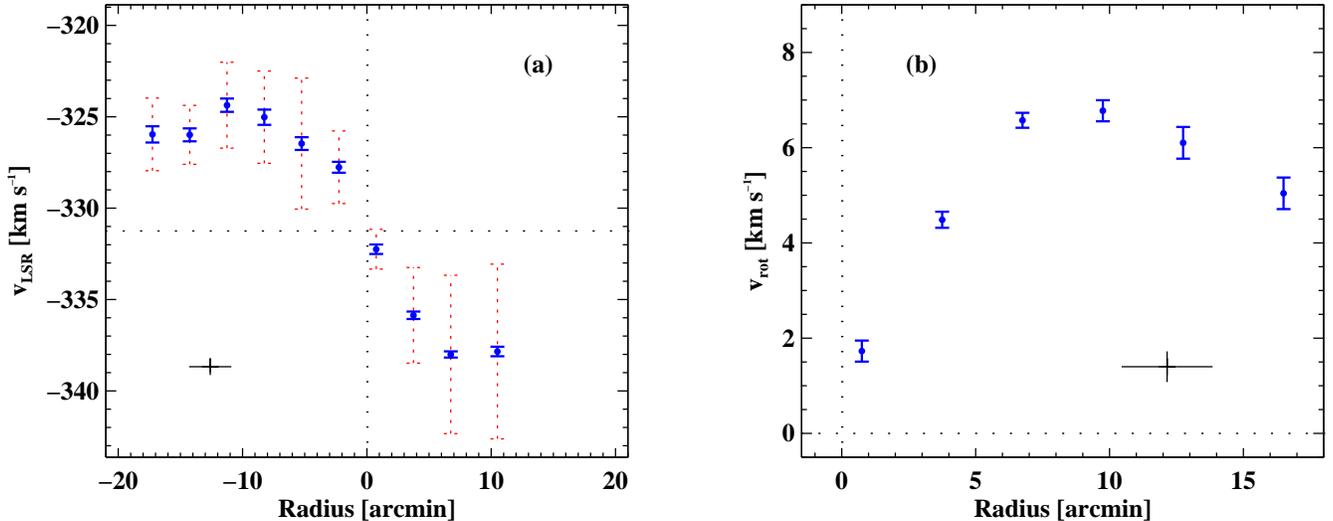}
\caption{({\it a}) The rotation curve along the major axis of the HVC.
The point of maximum symmetry is shown by the dotted lines and occurs
at $(\alpha,\delta)_{2000} = (01^{\rm\scriptstyle
h}05^{\rm\scriptstyle m}49^{\rm\scriptstyle s}, +21\degr 42\arcmin)$
with a systemic velocity of $-331.3\pm1.0\kms$. Blue error bars
represent the 1~$\sigma$ uncertainty in the velocity of the fitted
Gaussian; red dotted error bars represent the rms of the velocity
field perpendicular to the major axis.  Removing this rotation curve
from the velocity field yields a residual field with an rms of
3\kms. ({\it b}) The rotation curve folded about the point of maximum
symmetry. After a solid-body rise out to a radius of 10\arcmin, the
rotation curve begins to decrease. The error bars are the 1~$\sigma$
uncertainties of the weighted average of the Gaussian fits. The
crosses indicate the velocity and spatial resolution of our
measurements.
\label{fig:rc}}
\end{figure*}


\section{Results}
\label{sec:results}

Our map (Figure \ref{fig:map}) reveals considerable detail that was
not apparent in previous observations.  The \HI\ associated with LGS~3
is visible in the upper right (northwest) as the small red blob at
$(\alpha,\delta)_{2000} = (01^{\rm\scriptstyle h}03^{\rm\scriptstyle
m}54^{\rm\scriptstyle s}, +21\degr 53\arcmin)$.  The emission is
marginally resolved spatially and has an intensity-weighted mean LSR
velocity of $-287\kms$.  We measure an integrated \HI\ flux of $2.3
\pm 0.5$~Jy\kms, in agreement with the value of $2.7 \pm 0.2$~Jy\kms\
measured by \citet{youngl97}. For a distance of
700~kpc,\footnote{Three previous authors have measured distances to
LGS~3 of very close to 800~kpc: 810 \citep{lee95}; 770
\citep{apariciogb97}; 830 \citep{mould97}. However, in a recent paper
based on HST data, \citet{millerdlkh01} derive a distance of $620 \pm
20$~kpc.  We adopt an intermediate distance of 700~kpc for all
calculations.} the \HI\ mass of LGS~3 is $2.6 \times 10^5\ M_{\odot}$.
The extended emission seen previously
\citep{christiant83,hulsboschw88,blitzr00} now appears as a large,
double-lobed \HI\ cloud in the center of the map with a systemic
velocity of $-331.3 \pm 1.0\kms$, an integrated flux of $24 \pm
8$~Jy\kms, and a mean linewidth of 24\kms.  This cloud is completely
separate from LGS~3, located 30\arcmin\ away with a velocity
difference of $-45\kms$.  If the cloud is at the distance of LGS~3, it
contains 10 times as much \HI, making it unlikely that it originated
in LGS~3. Therefore, this cloud qualifies as an HVC.

Although LGS~3 and the HVC may not share a common origin, their
apparent proximity demands that we investigate the possibility of an
interaction betweeen them.  To the east of the HVC, at
$(\alpha,\delta)_{2000} = (01^{\rm\scriptstyle h}08^{\rm\scriptstyle
m}56^{\rm\scriptstyle s}, +21\degr 53\arcmin)$, a faint vertical strip
of \HI\ extends for $\sim$20\arcmin\ along the edge of the map. A
similar feature is also visible running horizontally in the southwest
at $(\alpha,\delta)_{2000} = (01^{\rm\scriptstyle
h}05^{\rm\scriptstyle m}00^{\rm\scriptstyle s},+21\degr 13\arcmin)$.
The symmetrical placement of this gas relative to the line connecting
LGS~3 and the HVC is suggestive of a tidal interaction, with the two
faint clouds representing leading and trailing tidal arms.  We
consider this idea further in \S 4.

The most striking feature of the HVC is the velocity gradient across
it.  We examined the gradient by averaging spectra perpendicular to
the major axis.  We fitted a Gaussian profile to the averaged spectrum
at each point along the major axis to create a rotation curve, which
is displayed in Figure \ref{fig:rc}{\it a}. The rotation curve
contains a roughly linear gradient over its central 20\arcmin\ and
then turns over.  The turnover and the overall symmetry of the
rotation curve suggest that the HVC is a single, rotating,
gravitationally-bound object, and is not composed of two physically
distinct clouds that happen to coincide along the line of sight.  In
order to measure a symmetric rotation curve from two separate clouds,
they would have to: have the same extent; have the same large aspect
ratio; be aligned along their major axes; have their velocity fields
vary with radius in just such a way as to mimic the appearance of a
single rotating cloud.  We consider this, although not impossible,
very unlikely.  We further point out that if the HVC were composed of
two overlapping clouds then the linewidths would be largest at the
center of the HVC.  Since we observe exactly the opposite, we conclude
that the HVC must be a single cloud.

We can therefore use the rotation curve to compare the dynamical mass
of the HVC to its luminous mass, which we assume consists only of \HI\
and a cosmic abundance of He, such that $\mlum=1.3\mhi$. The dynamical
mass (\mdyn) is the mass required to account for the rotational
velocity of the HVC at the last measured point (see Figure
\ref{fig:rc}{\it b}).  The \HI\ mass scales as the square of the
distance to the HVC, $\mhi = 5.5 \times 10^{6} (d/{\rm Mpc})^{2}\
M_{\odot}$, which is $2.7 \times 10^{6}\ M_{\odot}$ for a distance of
700~kpc.  The dynamical mass, however, scales linearly with distance,
$\mdyn = 2.8 \times 10^{7} (d/{\rm Mpc})\ M_{\odot}$, yielding
$2.0\times10^7 M_{\odot}$ at 700 kpc.  (Note that we assume that the
rotation is seen edge-on; if this is not the case, then the actual
dynamical mass will be larger by a factor of $1/\sin^2 i$.)
Therefore, if the HVC is indeed self-gravitating, as is suggested by
its rotation curve, and it is located at the distance of LGS~3, {\em
it must be composed mostly of dark matter} (82\%).  Furthermore, if
the cloud is assumed to be any closer than LGS~3 (and is
self-gravitating), its dynamical-mass-to-luminous-mass ratio
($\mdyn/\mlum$) must be even larger; if the cloud is further than
LGS~3, it remains dark-matter dominated out to a distance of 1.9~Mpc.
Since the darkest known galaxy-sized objects have $\mdyn/\mlum
\lesssim 100$ \citep[see][and references therein]{mateo98}, we can
place a firm lower limit on the distance to the HVC of 39~kpc.


\section{Discussion}
\label{sec:discussion}

Is the HVC indeed interacting with LGS~3?  We can approach this
question by considering the likelihood that these two objects are
completely unrelated.  \citet{putmanetal02} find that compact HVCs
cover $\lesssim$1\% of the southern sky.  \citet*{deheijbb02b} use an
automated analysis of the LDS to measure a similar covering fraction
for $\delta \gtrsim -30\degr$.  There are indications in the
\citet*{deheijbb02b} catalog of a factor of $\sim$2 overdensity of
compact HVCs within 20\degr\ of LGS~3, which is not surprising because
M31 and the Local Group barycenter are located in the same direction
as LGS~3.

We estimate the probability of a spatial coincidence between a dwarf
galaxy and a compact HVC by considering the number of compact HVCs and
dwarf galaxies near LGS~3.  For example, within 20\degr\ of LGS~3
there are 5 dwarf spheroidal galaxies and 14 compact HVCs from the
\citet*{deheijbb02b} catalog.  The HVC discussed in this paper is
located $\sim$30\arcmin\ away from LGS~3, so we take a circle of
radius 30\arcmin\ around each of the nearby dwarfs.  Now, the
probability that at least one of the compact HVCs lies within one of
these circles is $p \approx N_{\rm \scriptscriptstyle
CHVC}\Omega_{\rm\scriptscriptstyle dwarfs}/\Omega_{20}$, where
$N_{\rm\scriptscriptstyle CHVC}$ is the number of compact HVCs
contained in the region, $\Omega_{\rm\scriptscriptstyle dwarfs}$ is
the solid angle subtended by the circles around the five dwarf
galaxies, and $\Omega_{20}$ is the solid angle subtended by the region
within 20\degr\ of LGS~3.  This probability is 4.4\%.  If we increase
the region under consideration to within 60\degr\ of LGS~3, there are
10 dwarf spheroidals (including the small dwarf irregulars WLM and
Pegasus) and 46 compact HVCs, and the probability of a chance
coincidence is 3.5\%.  Including the probability of the velocities
coinciding within 50\kms\ will make the overall probability of such a
configuration even lower.  However, since these values are not
negligibly small, we cannot dismiss the possibility of a chance
superposition entirely.  Nevertheless, we conclude the HVC is probably
at the same distance as LGS~3.

Another piece of evidence that the HVC is not associated with the
Milky Way is its extremely large velocity with respect to the Galactic
Standard of Rest (GSR) of $\vgsr = -200\kms$.  An object originally
associated with the Galaxy (e.g., in a Galactic fountain) would have
difficulty acquiring such a high velocity.  This velocity also renders
a distance beyond the Local Group implausible.

Could this object be another dwarf galaxy interacting with LGS~3,
rather than an HVC?  \citet{simonb02} searched most of the compact
HVCs in the northern hemisphere and concluded that they do not contain
stellar counterparts similar to the known Local Group dwarf galaxies.
There is no obvious counterpart to this HVC on the Second Palomar
Observatory Sky Survey\footnote{The Second Palomar Observatory Sky
Survey (POSS-II) was made by the California Institute of Technology
with funds from the National Science Foundation, the National
Geographic Society, the Sloan Foundation, the Samuel Oschin
Foundation, and the Eastman Kodak Corporation.} plates.  We used the
techniques of \citet{simonb02} to search more carefully and found a
very faint low-surface brightness feature nearby.  However, follow-up
imaging with the Lick 3~m telescope revealed that there is no distant
stellar population here.  Therefore this cloud, like all other compact
HVCs, appears to be a pure gas cloud.

Based on our data, we suggest that the faint features to the east and
south of the HVC are remnants of a tidal interaction between the HVC
and LGS~3. We now consider the consequences of such a situation. The
total mass of LGS~3 (derived from its central velocity dispersion) is
$1.3 \times 10^{7}\ M_{\odot}$ \citep{mateo98}.  Hence, its tidal
field becomes comparable to the surface gravity of the HVC if the HVC
orbit has a minimum approach of $\sim$4~kpc. If the faint features are
tidal in origin, the orbit of the HVC is strongly constrained. Since
the gravity of LGS~3 is not enough to bind the system---the escape
velocity of LGS~3 is $\sim$19\kms, while the relative velocity
between LGS~3 and the HVC is $\gtrsim$50\kms---the encounter between
them must be a one-time event.  The existence of the tidal tails
implies that they have already made their closest approach.  Given
their relative velocity and their projected separation of 5.8~kpc,
such an approach would have taken place $\sim\!\!10^{8}$ years ago.


\section{Conclusions}
\label{sec:conclusions}

We have presented high-resolution \HI\ observations of a 2.4 deg$^{2}$
area including the Local Group dwarf galaxy LGS~3 and a previously
unresolved cloud of gas adjacent to the galaxy.  Our data show that
the \HI\ cloud is 30\arcmin\ away from the galaxy, with a velocity
difference of $-45\kms$, and that they do not appear to be connected.
If they are at the same distance, the cloud contains 10 times as much
\HI\ and twice as much total mass as LGS~3. Optical imaging of the
cloud revealed no stellar counterparts.  We therefore argue that the
cloud is an HVC, and should not be considered part of the \HI\
component of LGS~3.  However, the \HI\ morphology does suggest that an
interaction is occurring.

We propose that the faint thin strips of gas on either side of the HVC
are tidal arms produced by a close encounter with LGS~3.  We note that
the probability of a line-of-sight coincidence between the two
objects, if they are at different distances, is $\sim$4\%. We
therefore suggest that the most likely interpretation of this system
is that LGS~3 and the HVC are at the same distance and have recently
undergone a tidal interaction.

We also examine the rotation curve of the HVC, which exhibits a linear
increase with radius out to 10\arcmin\ and then begins to decrease.
Under the assumption that the HVC is self-gravitating, which is
supported by the rotation curve's symmetry and turnover at large
radii, we use the dynamical mass and the \HI\ mass to derive a lower
distance limit for the HVC.  For $\mdyn/\mlum$ to be $\lesssim 100$,
the HVC must be at least 39~kpc away if it is self-gravitating. The
extremely negative \vgsr\ of the HVC constrains it to be within the
Local Group. This HVC is thus unique in that there is evidence that it
is both dark-matter dominated and physically associated with an object
at a known distance.


\acknowledgments We thank Phil Perillat, Karen O'Neil, Mike Nolan, \&
Arun Venkataraman for their support during our observations and Carl
Heiles for his indispensable guidance. This work was supported in part
by NSF grant AST 99-81308.


\end{document}